# Cut-Matching Games on Directed Graphs

Anand Louis
Georgia Tech, Atlanta
anand.louis@cc.gatech.edu

November 1, 2018

**Abstract**

We give $O(\log^2 n)$-approximation algorithm based on the cut-matching framework of [10, 13, 14] for the computing the sparsest cut on directed graphs. Our algorithm uses only $O(\log^2 n)$ single commodity max-flow computations and thus breaks the multicommodity-flow barrier for computing the sparsest cut on directed graphs.

## 1 Introduction

The Directed Sparsest Cut problem (henceforth refered to as DSC ) is the following: Given a graph $G = (V, E)$, find a partition $(S, \bar{S})$ of $V$ which has the minimum directed edge expansion. The directed edge expansion of a cut $(S, \bar{S})$ is defined as $\frac{|\delta_E^{out}(S)|}{min\{|S|,|\bar{S}|\}}$ [1] . This problem is known to be NP-hard. Interest in this problem and its undirectd version derives both from its numerous practical applications such as image segmentation, VLSI layout and clustering (see the survey of Shmoys [16]), and from its theoretical connections to spectral methods (see [7]), linear/semidefinite programming and metric embeddings. In this paper, we provide fast algorithms for this problem using the framework proposed by Khandekar et. al. in [10] (KRV) and subsequently studied by Orecchia et. al. [13] and Orecchia et. al. [14] (OSVV). This framework reduces the NP-hard DSC problem to the computation of a poly-logarithmic number of single-commodity max-flows while, at the same time, keeping the approximation factor poly-logarithmic.

**The Cut-Matching Game framework** : KRV and OSVV studied the Cut-Matching game on undirected graphs. At the heart of the results of KRV and OSVV lies a two person game, which starts with an empty graph on $n$ vertices (assume $n$ is even): in each round, the CUTPLAYER chooses a bisection $(S, \bar{S})$ of the vertices and in response the MATCHINGPLAYER chooses a perfect matching between $(S, \bar{S})$. The game ends when the (multi)-graph consisting of the multi-set union of the perfect matchings has edge expansion at least $\alpha$, where $\alpha$ is a parameter of the game. The goal of the CUTPLAYER is to minimize the number of rounds of play, while the MATCHINGPLAYER tries to draw the game out for as many rounds as possible. Any strategy for the CUTPLAYER that guarantees termination in $t$ rounds yields a $O(t/\alpha)$-approximation algorithm for Sparsest Cut. Moreover the resulting algorithm runs in time $O(t(T_c + T_f))$ where $T_c$ is the time to implement the cut player and $T_f$ is the running time of a single commodity max-flow computation on a $n$-vertex graph. KRV gave a cut-player strategy that acheives $t = O(\log^2 n)$ and $\alpha = \Omega(1)$ , thereby obtaining an $O(\log^2 n)$ approximation algorithm for Sparsest Cut problem on undirected graphs. OSVV gave a cut-player strategy that acheives $t = O(\log^2 n)$ and $\alpha = \Omega(\log n)$ thereby obtaining a $O(\log n)$ approximation algorithm. In both these strategies the running time is dominated by the polylogarithmic number of max-flow computations.

**Related Work** Arora et. al. [6] gave a $O(\sqrt{\log n})$-approximation algorithm for the undirected sparsest cut problem. Their algorithm was based on semidefinite programming. The paper also introduced the framework of expander flows. The basic idea here is that the sparsity of the given graph $G$ can be closely approximated by finding an expander that can be embedded into $G$ with minimum congestion. Arora et. al. [4] gave an efficient multicommodity flow based implementation using the expander flow framework. They achieved running time $\tilde{O}(n^2)$ [2] for an $O(\sqrt{\log n})$ approximation. Breaking the multicommodity flow barrier to get a $O(\sqrt{\log n})$ approximation remained an open problem for many years till it was finally resolved by Sherman [15]. They gave an algorithm based on Arora et. al.'s framework, but used single-commodity max-flow computations between a polylogarithmic number of carefully chosen vertices instead of a multicommodity max-flow computation. On a different direction of study, Arora and Kale [5] gave an algorithm that achieves an approximation ratio of $O(\log n)$ while still running in time dominated by poly-logarithmic single commodity max-flow computations. Their algorithm worked in a more general framework for designing primal-dual algorithms for SDPs.

---

[1]$\delta_A^{out}(S) = \{(u,v) \in A | u \in S \ \& v \notin S \}$, $\delta_A^{in}(S) = \{(u,v) \in A | u \notin S \ \& v \in S \}$
[2]$\tilde{O}(f)$ denotes $O(f \text{ polylog } f)$

For DSC, Agarwal et. al. [2] gave a semidefinite programming based $O(\sqrt{\log n})$-approximation algorithm. A generalized version of this problem has been studied by Leighton et. al. [11] and by Agarwal et. al. [1]. Fast algorithms for single commodity max-flow computations have been studied in [8, 12].

**This work :** We propose a cut-player strategy for cut-matching games on directed graphs. We show that our cut-player strategy leads to an $O(log^2 n)$ approximation algorithm for the DSC. Our algorithm uses only $O(\log^2 n)$ single commodity max-flow computations. To the best of our knowledge this is the first study of cut-matching games on directed graphs, and the first polylogarithmic approximation algorithm for this problem which breaks the multicommodity flow barrier.

Our CUTPLAYER strategy is quite similar to that of KRV. Given a set of matchings, their cut player starts with a suitably chosen initial distribution on the vertices, and compute the probability distribution that results from running a random walk on the graph consisting of the union of the matchings. It then sorts the vertices according to their final probabilities and outputs the median cut, say $(S, \bar{S})$. Their matching-player computes a cut perfect matching across the cut. The key to obtaining an approximation algorithm from a cut-matching game is to ensure that the matching player produces a matching that is embeddable in the input graph. In the directed setting, this matching-player strategy clearly fails as outputting a directed matching from $S$ to $\bar{S}$ (or from $\bar{S}$ to $S$) clearly will not suffice if we want to union of matchings to be a (directed) expander. Trying to mirror the KRV analysis to the directed setting, one would want the MATCHINGPLAYER to find a symmetric matching [3]. But the input graph need not contain a symmetric matching. We introduce a relaxed notion of perfect matchings in directed graphs and show how to compute them using single commodity max-flows. The analysis of our cut-matching game is similar to that of KRV and OSVV in that it makes use of a potential function related to the mixing of the walk on the current union of matchings: However, bounding the change in potential function is now much trickier.

## 2 Review of the Cut-Matching Game framework

**Certifying expansion**. To certify that a given graph $G$ has no sparse cut, one could use the expander flow formalism of Arora et. al. [6]. This consists of constructing a graph $H$ of known expansion on $n$ vertices and embedding it as a flow in $G$ such that the flow routed through each edge in $G$ is at most 1. This would certify that the expansion of $H$ is a lower bound on the expansion of $G$. The seminal result of Leighton and Rao [11] may be viewed as a multi-commodity flow based algorithm to embed the scaled complete graph in any $n$-vertex graph $G$ or produce a sparse cut in $G$. The core of the algorithms in [6, 4, 10, 5, 14] is based on this expander flow formulation.

**The players**. The game consisting of 2 players is played on the input graph $G = (V, E)$. The CUTPLAYER takes as input a series of perfect matchings on the vertex set $V$ and out puts a bisection $(S, \bar{S})$ of the vertex set. The MATCHINGPLAYER takes as input this bisection and either outputs a perfect matching across it which is embeddable[4] in $G$ or it outputs a sparse cut to prove that no matching across the bisection exists. We formally present the algorithm based on the cut-matching game in Figure 1 (We assume that the CUTPLAYER and the MATCHINGPLAYER are oracles. In Section 3 we show how to implement them):

---

[3] A matching $M$ on a directed graph $G$ is a symmetric matching if $(u, v) \in M \implies (v, u) \in M$.

[4] A matching $M$ on the vertex set $V$ is said to be embeddable in a graph $G = (V, E)$ if $G$ is able to simultaneously support a unit flow from vertex $u$ to vertex $v$ $\forall (u, v) \in M$

1. Input graph $G = (V, E)$, $\alpha$
    (a) $t = 0$, $M_0 = \phi$
    (b) CutPlayer : Let $(S, \bar{S})$ be the cut returned by CutPlayer ($\{M_1, \ldots, M_t\}$)
    (c) MatchingPlayer :
        i. If MatchingPlayer $(S, \bar{S})$ successfully returns a matching $M$ between $S$ and $\bar{S}$ then
            A. $M_{t+1} := M$
            B. $t := t + 1$
        ii. If MatchingPlayer $(S, \bar{S})$ returns a cut $(C, \bar{C})$ then output $(C, \bar{C})$ and End.
    (d) If $\cup_{i=1}^{t} M_i$ is an $\alpha/2$-expander, then End.

Figure 1: The algorithm for sparsest cut based on the cut-matching game

## 3 The Algorithm

We say that a set of edges $M$ is a directed matching on the set of vertices $V$ if $|\delta_M^{in}(v)|, |\delta_M^{out}(v)| \leq 1$ $\forall v \in V$. The directed matching is a *perfect* directed matching if $|\delta_M^{in}(v)| = |\delta_M^{out}(v)| = 1$ $\forall v \in V$. For the rest of the paper, matching and perfect matching refer to directed matching and perfect directed matching respectively.

At a high level our algorithm may be viewed as iteratively building an expander $H$ that embeds in the graph $G$ with congestion $O(\log^2 n/\alpha)$. Let $\{M_1, \ldots, M_t\}$ be a sequence of perfect matchings. We define a $t$-step random walk associated with this sequence of matchings as follows : in the $i^{th}$ step, the particle stays put with probability $1/2$ and traverses the out-edge from the current location in $M_i$ with probability $1/2$. The sequence of matchings $\{M_1, \ldots, M_t\}$ is mixing if for any starting position, the probability of the particle reaching any vertex $v$ is atleast $1/2n$. Observe that a mixing sequence of matchings forms a graph with directed edge expansion $1/2$ (Lemma A.1).

As in KRV and OSVV, we use a potential function defined on $\{M_1, \ldots, M_t\}$ to measure how far from uniform the resulting distribution of the associated random walk is when starting from a random vertex. Formally, $\psi(t) = \sum_{i,j}(P_{ij}(t) - 1/n)^2$ where $P_{ij}(t)$ is the probability that a particle starting at $j$ reaches $i$ in the random walk associated with $\{M_1, \ldots, M_t\}$. Observe that $\psi(0) = n - 1$ and $\psi(t) \leq 1/4n^2$ then the sequence $\{M_1, \ldots, M_t\}$ is mixing.

The Algorithm in Figure 1 starts with an empty sequence of matchings in $G$, and while the sequence is not mixing, it tries to find a new matching to add to the sequence. Our MatchingPlayer will try to find a matching that the matching is embeddable in $G$ with congestion $1/\alpha$ and the CutPlayer will ensure that any matching across the cut it outputs will reduce the potential by a factor of $(1 - 1/\log n)$. If the MatchingPlayer succeeds, then $O(\log^2 n)$ iterations will suffice to produce a mixing sequence of matchings. In this case, the union of the matchings which forms a 1/2-edge expander can be embedded in $G$ with congestion $O((\log^2 n)/\alpha)$. In case the MatchingPlayer does not succeed in finding a matching in some iteration, then the MatchingPlayer outputs a cut in $G$ with expansion at most $\alpha$.

We present our CutPlayer and MatchingPlayer in Figures 2 and 3 respectively. The CutPlayer runs in $\tilde{O}(n)$ and the MatchingPlayer runs in max-flow time. Note that the matching output by the MatchingPlayer is not a subset of edges of the input graph $G$, but a matching that is embeddable in $G$. We will show that the game in Figure 1 using these CutPlayer and MatchingPlayer oracles will prove the following theorem :

**Theorem 3.1.** *Given a graph $G = (V, E)$ and an $\alpha$, there exists an algorithm that*

– *either outputs a cut of expansion at most $\alpha$*

– *or proves that every cut has expansion at least $\alpha/\log^2 n$ by embedding in $G$ an $\alpha$-expander with congestion at most $O(\log^2 n)$.*

*Morover the algorithm can be implemented using $O(\log^2 n)$ single-commodity max-flow computations.*

By doing a binary search on $\alpha$ we get an $O(\log^2 n)$-approximation algorithm for DSC :

**Corollary 3.2.** *There exists an $O(\log^2 n)$-approximation algorithm for* DSC *whose running time is dominated by a polylogarithmic number of single-commodity max-flow computations.*

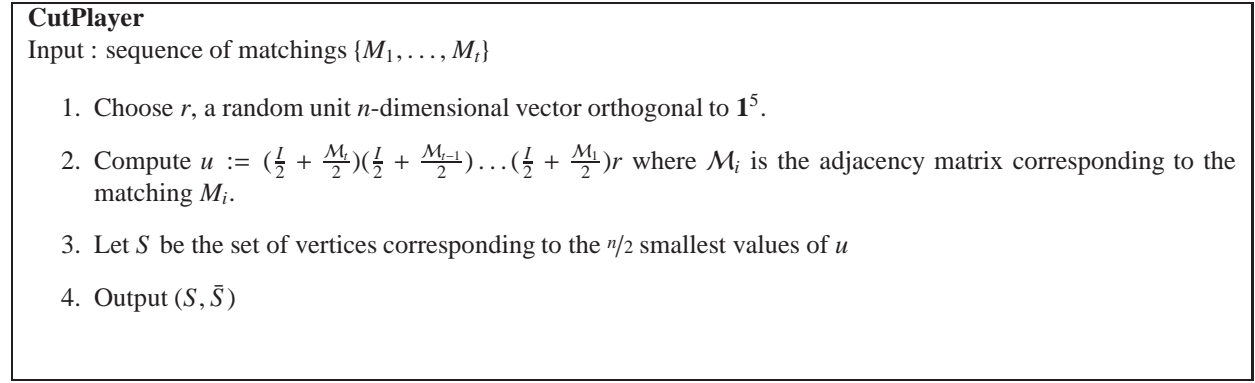

Figure 2: CutPlayer

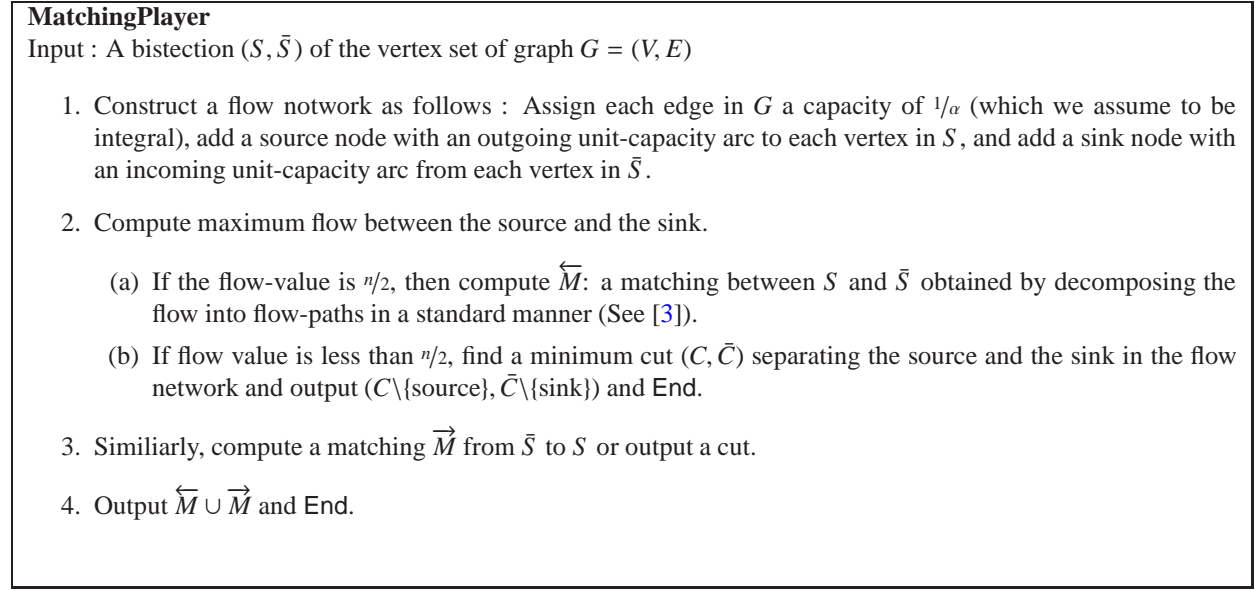

Figure 3: MatchingPlayer

## 4 Analysis

### 4.1 Analysis of the CutPlayer

Recall that $P_{ij}(t)$ is the probability of going from $j$ to $i$ in the natural random walk associated with $\{M_1, \ldots, M_t\}$. We define the vector $P_i(t) \stackrel{\text{def}}{=} (P_{i1}(t), \ldots, P_{in}(t))$ to denote the probability of ending up at $i$ starting from each vertex. Observe that the entries in $P_i(t)$ sum up to 1. This follows by induction as if $(i, j) \in M_t$ then $P_j(t) = (P_i(t-1) + P_j(t-1))/2$ (see Equation 1) and $\mathbf{1}^T . P_i(0) = 1 \ \forall i \in V$.

Note that the vector $u$ produced by CutPlayer on $M_1, \ldots, M_t$ is the projection of the vectors $P_i(t)$ onto the randomly chosen vector $r \perp \mathbf{1}$, i.e., $u_i = P_i(t).r$. The CutPlayer partitions the vertices into two sets according to whether the corresponding $u_i$s are large or small. Thus in any matching respecting this partition, if vertices $i$ and $j$ are matched, $|u_i - u_j|$ will tend to be large. Note that $u_i - u_j$ is the projection of $P_i(t) - P_j(t)$ on a random vector $r$. Since the vector $P_i(t) - P_j(t)$ lies in the $(n-1)$-dimensional space orthogonal to 1 and $r$ is a unit vector chosen uniformly at random from this space, we have $\mathbf{E}[|u_i - u_j|^2] = \|P_i(t) - P_j(t)\|^2/(n-1)$.

We make use of the following 3 lemmas :

**Lemma 4.1.** *The reduction in potential for a perfect matching $M$ is $\frac{1}{4} \sum_{(i,j)\in M} \|P_i(t) - P_j(t)\|^2$*

**Lemma 4.2.** *Whp, for all pairs $i, j$ we have $\|P_i(t) - P_j(t)\|^2 \geq \frac{n-1}{C \log n} \|u_i - u_j\|_1^2$*

**Lemma 4.3.** *For a perfect matching $M$ output by MatchingPlayer across a partition found by CutPlayer, we have*
$(n-1) \mathbf{E}[\sum_{(i,j)\in M} |u_i - u_j|^2] \geq 2\psi(t)$

The inequality in Lemma 4.2 holds with probability at least $1 - n^{-\Omega(1)}$. Clearly, these three lemmas together imply that the expected reduction in the potential in any iteration is $(1/\log n)$ fraction of the current potential. This implies that in $O(\log^2 n)$ iterations the potential drops below $1/4n^2$ with high probability and that union of matchings forms an expander. We now prove these lemmas.

*Proof. of Lemma 4.1*: For a particle starting at vertex $k$ to reach vertex $i$ in $t$ steps by following the natural random walk defined on the the sequence $\{M_1, \ldots, M_t\}$, it can either reach $i$ in $t-1$ steps and stay there in the next round with probability $1/2$, or it can reach a vertex $j$ such that $(j, i) \in M_t$ in $t-1$ steps and traverse $(j, i)$ in the $t^{th}$ step with probability $1/2$.

Therefore, $p_{ik}(t) = (p_{ik}(t-1) + p_{ij}(t-1))/2$. Generalizing this, we get

$$P_i(t) = \frac{P_i(t-1) + P_j(t-1)}{2} \tag{1}$$

Recall that $\psi(t) = \sum_{i\in V} \|P_i(t) - \mathbf{1}/n\|^2$. On adding a perfect matching $M$ across the cut produced by the CutPlayer the decrease in potential $\Delta\psi(t)$ is

$$\begin{aligned}
\Delta\psi(t) &= \sum_{i\in V} \|P_i(t) - \mathbf{1}/n\|^2 - \sum_{i\in V} \|P_i(t+1) - \mathbf{1}/n\|^2 \\
&= \sum_{i\in V} \|P_i(t) - \mathbf{1}/n\|^2 - \sum_{(i,j)\in M} \|\frac{P_j(t)+P_i(t)}{2} - \mathbf{1}/n\|^2 \\
&= \sum_{i\in V} \|P_i\|^2 - \sum_{(i,j)\in M} \|\frac{P_i+P_j}{2}\|^2 \qquad \text{(where } P_i = P_i(t) - \mathbf{1}/n\text{)} \\
&= \frac{\sum_{i\in V} \|P_i\|^2}{2} - \frac{\sum_{(i,j)\in M} P_i.P_j}{2} \\
&= \frac{\sum_{(i,j)\in M} \|P_i - P_j\|^2}{4} \\
&= \frac{\sum_{(i,j)\in M} \|P_i(t) - P_j(t)\|^2}{4}
\end{aligned}$$

□

*Proof. of Lemma 4.2*: Observe that $u_i - u_j$ is the projectiong of $P_i(t) - P_j(t)$ onto $r$. The proof for this lemma follows from the gaussian behavior of projections and has been proved in [10].

□

*Proof. of Lemma 4.3*: Let $\overrightarrow{M}$ and $\overleftarrow{M}$ be the 2 components of $M$ (see MatchingPlayer in Figure 3). Let $(S, \bar{S})$ be the bisection found by CutPlayer. Recall that $S$ is the set of $n/2$ vertices $i$ with smallest values of $u_i$. Let $\eta$ be a real number so that $u_i \leq \eta \leq u_j$ for any $i \in S$ and $j \in \bar{S}$. We then have :

$$\begin{aligned}
\sum_{(i,j)\in \vec{M}} |u_i - u_j|^2 &\geq \sum_{(i,j)\in \vec{M}}((u_i - \eta)^2 + (\eta - u_j)^2) \\
&= \sum_{i\in V}(u_i - \eta)^2 \\
&= \sum_{i\in V} u_i^2 - 2\eta \sum_{i\in V} u_i + n\eta^2 \\
&= \sum_{i\in V} u_i^2 - 2\eta \sum_{i\in V} P_i(t).r + n\eta^2 & (u_i = P_i(t).r) \\
&= \sum_{i\in V} u_i^2 - 2\eta \mathbf{1}.r + n\eta^2 & (\sum_{i\in V} P_i(t) = \mathbf{1}) \\
&= \sum_{i\in V} u_i^2 - 0 + n\eta^2 & (\mathbf{1}^T.r = 0) \\
&\geq \sum_{i\in V} u_i^2
\end{aligned}$$

Similarly we get that $\sum_{(i,j)\in \overleftarrow{M}} |u_i - u_j|^2 \geq \sum_{i\in V} u_i^2$.

We make use the following well known lemma about the gaussian behavior of projections.

**Lemma 4.4.** *If $v$ is a vector of length $l$ in $\mathbf{R}^d$ and $r$ is a random unit vector in $\mathbf{R}^d$ then $\mathbf{E}[(v^T r)^2] = l^2/d$*

We use this lemma for vectors $(P_i(t) - \mathbf{1}/n)$ which lie in the $(n - 1)$-dimensional space orthogonal to $\mathbf{1}$. Note that $u_i = P_i(t).r = (P_i(t) - \mathbf{1}/n).r$ denotes the projection of $(P_i(t) - \mathbf{1}/n)$ onto $r$. Since $r$ is a random unit vector from the space orthogonal to $\mathbf{1}$, we have $\mathbf{E}[u_i^2] = \|P_i(t) - \mathbf{1}/n\|^2/(n-1)$.

Hence we have

$$\begin{aligned}
\mathbf{E}[\sum_{(i,j)\in M} |u_i - u_j|^2] &\geq 2\mathbf{E}[\sum_{i\in V} u_i^2] \\
&= 2\frac{\sum_{i\in V} \|P_i(t) - \mathbf{1}/n\|^2}{n-1} \\
&= 2\frac{\psi(t)}{n-1}
\end{aligned}$$

□

Thus we have shown that (whp) the game will end in $O(\log^2 n)$ rounds.

## 4.2 Analysis of MATCHINGPLAYER

If the MATCHINGPLAYER is not able to find a perfect matching across the input bisection, then it outputs a cut. We now prove that that cut has expansion at most $\alpha$.

**Lemma 4.5.** *In the procedure MATCHINGPLAYER, if the max flow between the source and the sink is less than $n/2$, then the cut output by MATCHINGPLAYER has expansion at most $\alpha$,*

*Proof.* If the flow has value less than $n/2$, the minimum cut separating the source and the sink has capacity less than $n/2$. Let the number of edges in the cut incident to the source (resp. sink) be $n_s$ (resp. $n_t$). The remaining capacity of the cut is less than $n/2 - n_s - n_t$, and thus uses at most $\alpha(n/2 - n_s - n_t)$ edges in the original graph. Moreover, the cut consisting of edges in the graph separates at least $n/2 - n_s$ vertices in source-side from $n/2 - n_t$ vertices in sink-side. The expansion of this cut is at most $\alpha(n/2 - n_s - n_t)/min(n/2 - n_s, n/2 - n_t)$ which is at most $\alpha$.

□

## 5 Conclusions

The techniques introduced in this paper can be used to obtain polylogarithmic approximation algorithms running in (single commodity) max-flow time for some more problems on directed graphs like balanced separator problem, and some slight generalizations of the sparsest cut problem (the details will appear in the full version of the paper).

It would be intersting to see if one can obtain a $O(\sqrt{\log n})$-approximation algorithm for DSC problem that runs in single commodity max-flow time.

# A  Appendix

**Lemma A.1.** *A mixing sequence of matchings $\{M_1, \ldots, M_t\}$ forms a graph with edge expanion $1/2$.*

*Proof.* Consider the *timeline graph* $H = (V', E')$ defined on the vertex set $V * \{0, \ldots, t\}$ with arcs of the form $((i, j), (i, j+1))$ and $((i, j), (i', j+1))$ where $(i, i') \in M_{j+1}$. We also set the capacity of every graph as $1/2$. The random-walk as described above starting from a vertex $v \in V$ induces a flow in $H$ from $(x, 0)$ to the group $\{(1, t), \ldots, (n, t)\}$. If the walk mixes, then the random walk delivers atleast $1/2n$ unit of flow to each $(1, t), \ldots, (n, t)$. Using induction it can be shown that concurrent unit flows from $(1, 0), (2, 0), \ldots$ and $(n, 0)$ do not violate the edge capacities of $H$. Observe that each arc in the union of the matchings corresponds to exactly 1 arc in $H$. Thus, mapping the flow in $H$ to the union of matchings, we get that the latter is also able to support a concurrent unit-flow from each vertex such that each vertex is able to receive at least $1/2n$ unit flow from every other vertex. Therefore, for any cut $(S, \bar{S})$, the union of matchings support $|S|*(n-|S|)/2n \geq n/2$ units of flow across the cut (in both directions). Hence, every cut has expansion at least $1/2$.  □